\documentclass[conference]{IEEEtran}
\IEEEoverridecommandlockouts
\usepackage{cite}
\usepackage{amsmath,amssymb,amsfonts}
\usepackage{algorithmic}
\usepackage{graphicx}
\usepackage{textcomp}
\usepackage{xcolor}
\usepackage{hyperref}
\usepackage{physics}

\usepackage{fancyhdr}

\def\BibTeX{{\rm B\kern-.05em{\sc i\kern-.025em b}\kern-.08em
    T\kern-.1667em\lower.7ex\hbox{E}\kern-.125emX}}

\begin{document}

\title{A Practical Framework for Assessing the Performance of Observable Estimation in Quantum Simulation

\thanks{This work was done in collaboration with the Quantum Economic Development Consortium (QED-C) and was performed under the auspices of the QED-C Technical Advisory Committee on Standards and Performance Metrics. The authors acknowledge many committee members for their input to and feedback on the project and this manuscript.}
}

\author{\IEEEauthorblockN{Siyuan Niu}
\IEEEauthorblockA{\textit{University of Central Florida} \\
Orlando, FL 32816, USA \\
\textit{Computational Research Division} \\
\textit{Lawrence Berkeley National Laboratory}\\
Berkeley, CA 94720, USA 
}

\and

\IEEEauthorblockN{Efekan K\"okc\"u}
\IEEEauthorblockA{\textit{Computational Research Division} \\
\textit{Lawrence Berkeley National Laboratory}\\
Berkeley, CA 94720, USA 
}

\and

\IEEEauthorblockN{Sonika Johri}
\IEEEauthorblockA{\textit{Coherent Computing Inc.} \\
Cupertino, CA 
}

\and

\IEEEauthorblockN{Anurag Ramesh}
\IEEEauthorblockA{\textit{Davidson School of Chemical Engineering} \\
\textit{Purdue University}\\
West Lafayette, IN 47907, USA 
}

\and

\IEEEauthorblockN{Avimita Chatterjee}
\IEEEauthorblockA{\textit{Computer Science \& Engineering} \\
\textit{Pennsylvania State University}\\
State College, PA 16801 USA 
}

\and

\IEEEauthorblockN{David E. Bernal Neira}
\IEEEauthorblockA{\textit{Davidson School of Chemical Engineering} \\
\textit{Purdue University}\\
West Lafayette, IN 47907, USA 
}

\and

\IEEEauthorblockN{Daan Camps}
\IEEEauthorblockA{\textit{National Energy Research Scientific Computing Center (NERSC)} \\
\textit{Lawrence Berkeley National Laboratory}\\
Berkeley, CA 94720, USA 
}

\and

\IEEEauthorblockN{Thomas Lubinski}
\IEEEauthorblockA{\textit{QED-C Technical Advisory Committee - Standards} \\
Arlington, VA 22209, USA \\
\textit{Quantum Circuits Inc.}\\
New Haven, CT 06511 
}
}

\maketitle


\begin{abstract}
Simulating dynamics of physical systems is a key application of quantum computing, with potential impact in fields such as condensed matter physics and quantum chemistry. However, current quantum algorithms for Hamiltonian simulation yield results that are inadequate for real use cases and suffer from lengthy execution times when implemented on near-term quantum hardware. In this work, we introduce a framework for evaluating the performance of quantum simulation algorithms, focusing on the computation of observables, such as energy expectation values.
Our framework provides end-to-end demonstrations of algorithmic optimizations that utilize Pauli term groups based on $k$-commutativity, generate customized Clifford measurement circuits, and implement weighted shot distribution strategies across these groups.
These demonstrations span multiple quantum execution environments, allowing us to identify critical factors influencing runtime and solution accuracy.
We integrate enhancements into the QED-C Application-Oriented Benchmark suite, utilizing problem instances from the open-source HamLib collection. Our results demonstrate a 27.1\% error reduction through Pauli grouping methods, with an additional 37.6\% improvement from the optimized shot distribution strategy. Our framework provides an essential tool for advancing quantum simulation performance using algorithmic optimization techniques, enabling systematic evaluation of improvements that could maximize near-term quantum computers' capabilities and advance practical quantum utility as hardware evolves.
\end{abstract}



\pagestyle{fancy}

\renewcommand{\headrulewidth}{0.0pt}
\lhead{}
\rhead{\thepage}

\renewcommand{\footrulewidth}{0.4pt}
\cfoot{}
\lfoot{A Practical Framework for Assessing the Performance of Observable Estimation in Quantum Simulation}
\rfoot{\today}
\vspace{2cm}


\section{Introduction}
\label{sec:introduction}

Simulating physical systems is one of the most promising areas for quantum computation. As quantum computing hardware and software advance rapidly, the scale and sophistication of algorithms that can be tested grow accordingly~\cite{brown20245}. New algorithmic techniques are regularly emerging in the search for quantum utility~\cite{preskill2018quantum,brooks2019}, yet today's noisy intermediate-scale quantum (NISQ) devices remain limited in their capabilities. With the promise of systems that implement quantum error correction, incremental improvements are expected in the coming years.
But how do we know how far along we are in making progress towards quantum utility?

Benchmarking provides a structured approach for assessing quantum computing performance across different levels—component, system, and application~\cite{proctor2024benchmarkingquantumcomputers}. It serves multiple purposes, from comparing hardware platforms to tracking algorithmic improvements~\cite{PhysRevA.77.012307, PhysRevLett.106.180504, Blume-Kohout2017-no, Cross_2019, Boixo_2018, wack_clops_2021, parekh2016benchmarking, chen2022veriqbench, li2023qasmbench, cornelissen2021scalable, tomesh2022supermarq, miessen2024benchmarking}. A key challenge in assessing quantum utility is determining whether a quantum computer can efficiently solve meaningful problems. As new algorithmic techniques are developed, benchmarking enables direct performance comparisons, helping to quantify advancements in solving real-world problems.

\begin{figure}
    \centering
    \includegraphics[ scale=0.45]{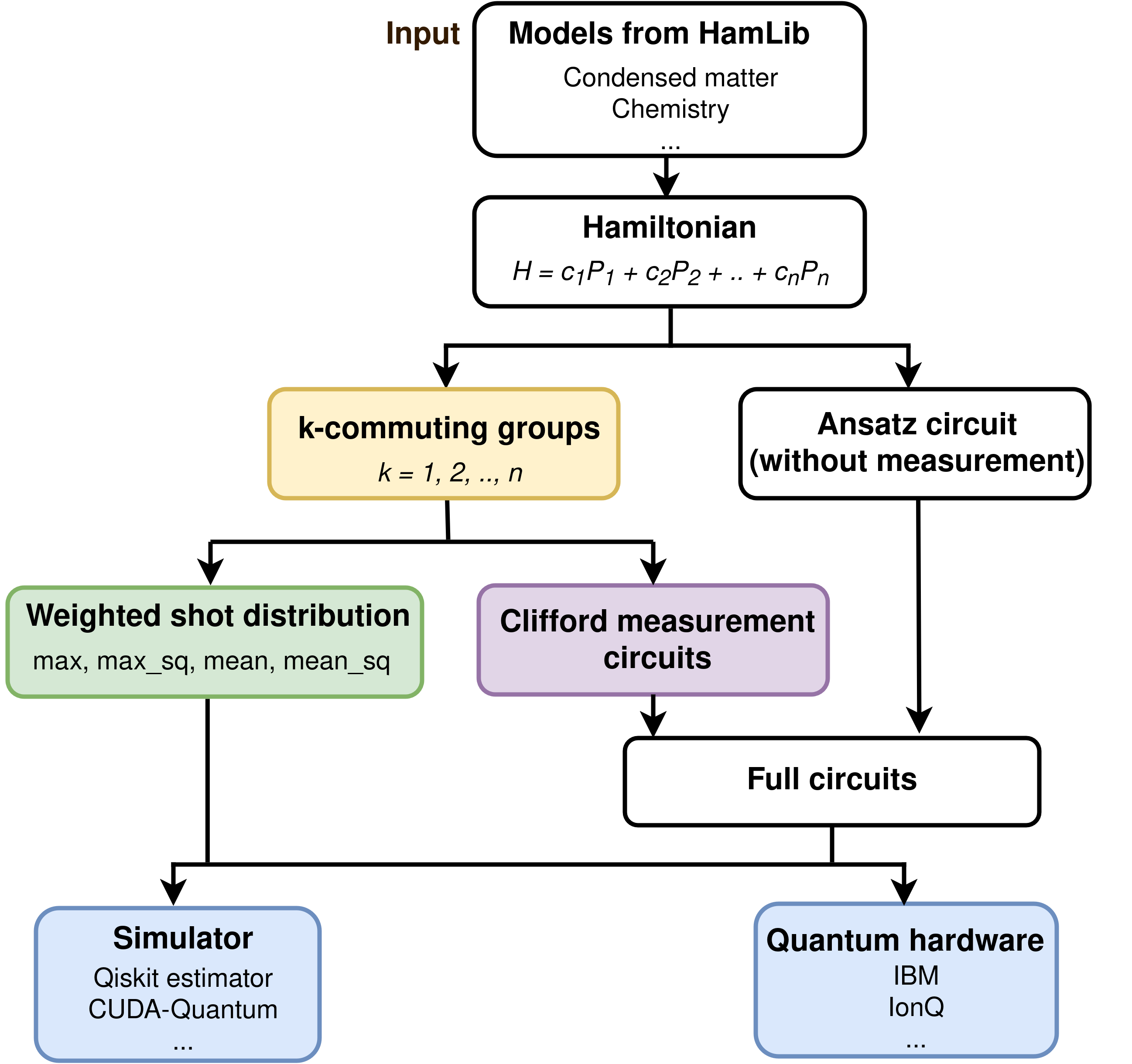}
    \caption{\textbf{Proposed framework for assessing quantum simulation performance.} These plots illustrate the structure of the proposed framework for evaluating the performance of Hamiltonian simulation under various proposed algorithmic optimization techniques. }
    \label{fig:1}
\end{figure}

Simulating physical systems on a quantum computer involves multiple interdependent components, each affecting overall performance. One of the most critical factors is minimizing the number of circuit executions and total shot count, as these directly impact both solution accuracy and execution time. As problem size increases, measurement variance grows, necessitating more shots to achieve statistical confidence. This challenge is particularly evident in algorithms such as the Variational Quantum Eigensolver (VQE), which, despite using shallow circuits, requires numerous executions to optimize a solution.

Here, we introduce novel techniques to improve the efficiency and accuracy of observable computation in quantum simulations and present an enhanced benchmarking framework that incorporates these techniques for performance evaluation. \autoref{fig:1} illustrates the framework, with the proposed methods highlighted in color blocks, along with their impact on the algorithm’s runtime. The introduced techniques include grouping Pauli terms in the Hamiltonian based on commutativity, constructing Clifford-based measurement circuits, and distributing shots to these circuits according to the coefficients of Pauli terms within the group. These methods reduce both circuit executions and shot count while preserving accuracy, addressing a key bottleneck in Hamiltonian simulation algorithms.

To support end-to-end assessment, we extend the existing QED-C Application-Oriented Benchmark Suite to evaluate performance across critical computational steps, including quantum evolution and observable estimation.  QED-C has a collection of quantum algorithms and applications that are helpful in evaluating the performance of quantum computing systems in addressing real-world problems. The suite constructs quantum computing solutions to problems derived from Hamiltonians extracted from the open-source HamLib collection~\cite{sawaya2023hamlib}. Our enhanced framework focuses on simulating quantum systems with flexible access to the measurement of any observables. It is designed to be highly configurable, allowing users to tune the setup and parameters of each proposed algorithmic optimization technique. This modularity also makes it compatible with other algorithmic approaches and adaptable to different quantum simulation workflows.

Using this extended framework, we demonstrate a 27.1\% reduction in observable estimation error compared with generalized estimator algorithms by applying a $k$-commuting grouping method for Pauli terms that construct the Hamiltonian along with a customized Clifford-based measurement circuit construction. Furthermore, by incorporating a weighted shot distribution algorithm, we achieve an additional 37.6\% average error reduction, further enhancing estimation accuracy. These algorithmic optimization techniques also reduce the number of shots required to reach a given level of accuracy, thereby improving overall runtime efficiency.

The framework is implemented in both Qiskit and CUDA Quantum, enabling access to high-quality noiseless simulations on GPUs to support algorithm development, while larger-scale quantum hardware continues to progress toward fault-tolerant quantum computing (FTQC). We also discuss how this environment can help identify the tipping point where tradeoffs between time and solution quality lead to practical quantum utility.

This paper is structured as follows.
Section~\ref{sec:background} provides background on Hamiltonian simulation, the HamLib library, and the QED-C’s application-oriented benchmarking suite.
In Section~\ref{sec:methods_for_estimating_observables}, we introduce various methods for observable estimation, including three existing techniques for comparison and our proposed algorithmic optimizations.
Section~\ref{sec:results-analysis} presents practical performance evaluations, focusing on improvements in observable estimation fidelity and execution runtime, and includes results from both simulation and real quantum hardware.
Finally, Section~\ref{sec:summary-and-conclusions} outlines future research directions and concludes with a summary of our findings.

\section{Background}
\label{sec:background}

This section provides background on the key concepts and tools underlying our work. We briefly review the Hamiltonian simulation and observable estimation, the HamLib collection of sample Hamiltonians, and the QED-C Application-Oriented Performance Benchmark framework, which we extend to assess simulation performance. While many detailed references exist on these topics, our aim here is to provide the necessary context for the work presented in this paper.


\subsection{Hamiltonian Simulation}
\label{subsec:hamiltonian-simulation}

In quantum computing, Hamiltonians play a central role in simulating physical systems, as they encode interactions and dynamics in fields such as chemistry, condensed matter physics, and materials science~\cite{von2021quantum,mcardle2020quantum,denner2023hybrid,savage2024quantum}.
A Hamiltonian is a mathematical construct of quantum mechanics, the operator $\hat{H}$, that determines the total energy of a physical system and precisely specifies its dynamics~\cite{Eisberg_Resnick_2017a}.
Computing the system's state as it evolves over time exposes its dynamics while computing the lower-energy eigenstates can reveal its steady-state properties. These tasks form the basis of two prominent quantum simulation applications: \textit{quantum dynamics simulation}, one of the simplest methods is the first-order Trotterization to approximate time evolution, and \textit{ground-state energy estimation}, frequently performed using the Variational Quantum Eigensolver (VQE) algorithm. Beyond physical simulations, specially constructed Hamiltonians are also used in quantum algorithms for combinatorial optimization problems, linear, and differential equation solvers~\cite{farhi2014quantum, abbas2023quantum,lloyd2010quantum,liu2021efficient,Lucas_Karp, nielsen2010quantum, cao2019quantum, kandala2017hardware, kirby2021variational}.

A Hamiltonian is typically expressed as a linear combination of Pauli operators, \( H = c_1 P_1 + c_2 P_2 + \dots + c_L P_L \), where \( P_i \) are tensor products of Pauli matrices. Algorithms for simulating time evolution generate circuits that consist of rotations with respect to the Pauli strings $P_i$. As an example, a first-order Trotter circuit with $r$ time steps for such a Hamiltonian can be constructed using the following expression with the associated error bound \cite{childs2021theory}:
\begin{align}
    e^{-itH} = \left( e^{-i\frac{t}{r}c_1 P_1}e^{-i\frac{t}{r}c_2 P_2} \dots e^{-i\frac{t}{r}c_L P_L}\right)^r + O\left( \frac{t^2}{r}\right)
\end{align}
Our framework is capable of generating such a time evolution circuit automatically upon the user's request.


The ground state of a Hamiltonian consisting of Pauli strings can be estimated via variational methods such as VQE, QAOA, and their variations \cite{farhi2014quantum, peruzzo2014variational, grimsley2019adaptive, tang2021qubit, romero2018strategies}. The majority of these methods rely on efficient measurement of the expectation value of the Hamiltonian operator on a variational state, i.e., $\bra{\psi(\vec{\theta})}H\ket{\psi(\vec{\theta})}$, where $\vec{\theta}$ are the parameters that are varied to minimize the expectation value. The most straightforward way of measuring the expectation value of the Hamiltonian $H = \sum_{i=1}^L c_i P_i$ is measuring the expectation values of the Pauli strings $P_i$ one by one, which requires $L$ circuits. For real-world Hamiltonians that may have hundreds of terms, the execution time can grow very large, thus moving some problems into a regime impractical for a quantum solution. To address this issue and reduce the required number of circuits, commuting Pauli strings $P_i$ can be grouped and measured together by mutually diagonalizing each observable in the group via a measurement basis  \cite{dalfavero2312k, yen2021cartan, sawaya2024nonclifforddiagonalizationmeasurementshot}. We implement the $k$-commutativity method given in \cite{dalfavero2312k} in our framework and discuss both the method and its effectiveness over different sets of Hamiltonians throughout this paper.

\subsection{HamLib: A Library of Hamiltonians}
\label{subsec:hamlib}

HamLib~\cite{sawaya2023hamlib} is a comprehensive dataset of quantum Hamiltonians, encompassing problem sizes ranging from 2 to 1000 qubits. The Hamiltonians are organized into several high-level categories. A major category includes condensed matter physics models, such as the transverse-field Ising model, the Heisenberg model, the Fermi-Hubbard model, and the Bose-Hubbard model. Another category contains chemistry Hamiltonians, incorporating curated or computed real-world parameters, with subcategories covering electronic structure and vibrational structure problems. Additionally, the dataset includes discrete-variable optimization problems, such as Max-K-Cut and the traveling salesperson problem. Finally, it features binary-variable optimization problems, including Max-K-SAT, Max-Cut, and QMaxCut.

A valuable feature of HamLib is that all problem instances have already been mapped to qubits, i.e., they are mapped to a Pauli representation of the form
$
\hat{H}_{\text{encoded}} = \sum_{i} c_i \bigotimes_k \sigma_{ik}
$,
where $\sigma_{ik}$ is a one-qubit Pauli or identity operator, i.e., $\sigma_{ik} \in \{I, X, Y, Z\}$, and $c_i$ is a real number. We select multiple Hamiltonians from this dataset, construct circuits based on them, and benchmark their performance. We systematically vary the methods by which the circuits are constructed, and the observables computed and assess the performance of each approach.

\subsection{QED-C Application-Oriented Benchmarks}
\label{sec:application_oriented_benchmarks}

\begin{figure}[t!]
\centering
    \includegraphics[width=0.76\columnwidth]{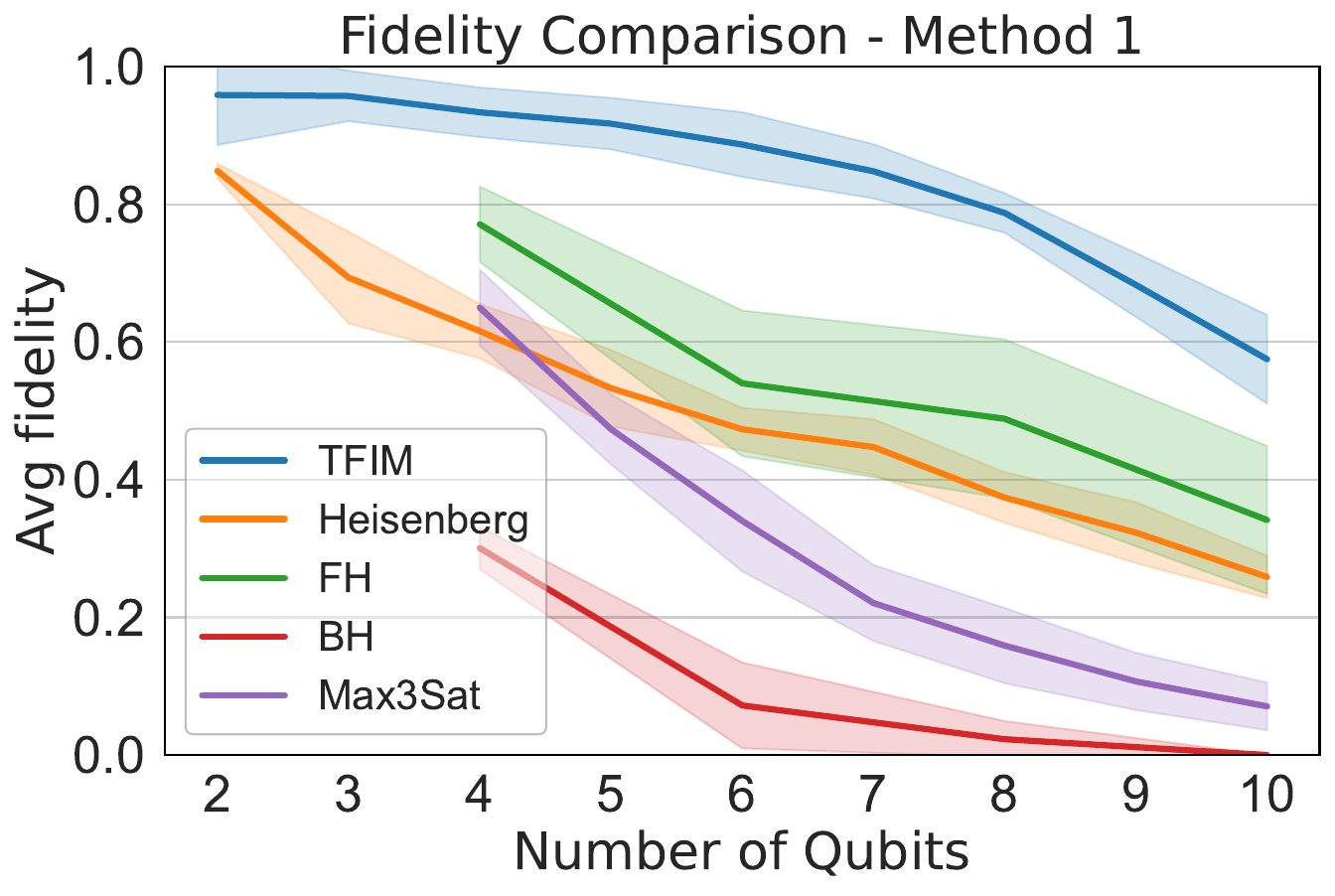}
    \caption{\textbf{QED-C Application-Oriented Hamiltonian Simulation Benchmark}
    The QED-C benchmark suite evaluates quantum applications across systems. This figure shows results for five HamLib Hamiltonians—TFIM, Heisenberg, Fermi-Hubbard, Bose-Hubbard, and Max3SAT—simulated using five-step Trotterized evolution (total time = 1.0). Circuits with 2 to 10 qubits were executed with 1000 shots on a classically implemented quantum simulator with quantum volume 2048 and compared to an ideal quantum simulator, with shaded regions indicating parameter-dependent variations in HamLib.
    }
    \label{fig:hamlib_fidelity_comparison}
\end{figure}

The Quantum Economic Development Consortium (QED-C) has developed an open-source benchmark suite that builds upon prior years of quantum benchmarking methodologies~\cite{PhysRevA.77.012307, PhysRevLett.106.180504, Blume-Kohout2017-no, Cross_2019, Boixo_2018, wack_clops_2021}. This suite enables comprehensive performance assessment across various quantum computing systems, including high-performance NVIDIA GPUs~\cite{lubinski2023_10061574,lubinski2023optimization, lubinski2024quantum}. It measures execution quality, runtime costs, and resource requirements for both single-circuit executions and iterative algorithms like QAOA~\cite{farhi2014quantum} and VQE~\cite{peruzzo2014variational}, using normalized Hellinger fidelity to assess circuit quality under noise~\cite{lubinski2023_10061574}. 

The benchmark suite includes support for assessing the performance of Hamiltonian simulation implementations, using problems from HamLib~\cite{chatterjee2024comprehensivecrossmodelframeworkbenchmarking}, as shown in \autoref{fig:hamlib_fidelity_comparison}. Several Hamiltonians are evolved using a 5-step Trotterized circuit, with measurement results compared against classical quantum simulator results using three different methods. The fidelity results from one of the methods, plotted against qubit count, show rapid degradation in quality, below 0.5 at only 6 qubits for some Hamiltonians.

The QED-C approach differs from system-level benchmarks such as Quantum Volume (QV)~\cite{cross2019validating, qiskit_measuring_quantum_volume, baldwin2022re, pelofske2022quantum}  and Volumetric Benchmarking (VB)~\cite{proctor2022measuring, blume2020volumetric, proctor2022establishing} by providing well-defined programs that yield application-specific performance metrics that can be adapted to various quantum hardware and simulators~\cite{lubinski2023_10061574, lubinski2023optimization, lubinski2023_10061574}.
The QED-C suite also reports execution time, which is particularly important for iterative algorithms where runtime overhead accumulates with multiple circuit executions.

Building upon the Hamiltonian simulation benchmarking framework, we extend our analysis to evaluate the performance of quantum algorithms that compute observable values from an ansatz or time-evolved quantum states and the associated Hamiltonians.

\section{Methods for Estimating Observables}
\label{sec:methods_for_estimating_observables}

Estimating observable values on a quantum computer requires executing multiple circuits that extend the base simulation with specific basis rotation operations and measurements corresponding to terms in the Hamiltonian. The accuracy of these computed values depends on both hardware noise and execution parameters, such as shot count, presenting important performance tradeoffs. Our framework evaluates different methods for observable computation across various implementation settings, collecting metrics on both execution time and result quality. A detailed performance analysis is presented in Section~\ref{sec:results-analysis}.

In theory, quantum computation promises scalability, but computing observables from evolved quantum states presents two practical challenges. First, the number of Hamiltonian terms typically grows polynomially (often quadratically) with problem size, increasing the number of required circuit executions and overall computational cost. Second, larger qubit counts demand more measurement shots to achieve statistical significance, further amplifying resource requirements. While prior studies have explored quantum Hamiltonian simulation and its resource demands~\cite{childs2021theory, low2023complexity, Moses_2023, dong2022quantum,gao2020benchmarking, saller2021benchmarking, yu2024qh9, baniasadi2018new}, many focus on specific problem instances or hardware architectures, often employing device-specific optimizations to maximize performance.

The expectation value of a Hamiltonian \( H \) for a quantum state \( |\psi\rangle \) is given by $\langle\psi|H|\psi\rangle$.  While classical computers can compute this directly for small systems, the exponential growth of the state space limits this approach unless special symmetries or constraints reduce the effective problem size. A quantum computer instead computes the expectation value in a scalable way by decomposing the Hamiltonian into a sum of weighted Pauli terms, $H = \sum_i c_i P_i$,  
allowing measurements to be performed term-by-term in a way that remains efficient even as the system size increases.

For each Pauli term in the Hamiltonian, a separate circuit is created by appending appropriate basis rotation gates to the quantum state preparation circuit. A total measurement shot budget is then allocated across these circuits. For each circuit, the assigned number of shots is executed to obtain a measurement outcome distribution. Expectation values are calculated by summing the measurement counts, weighted by \( \pm1 \) based on the parity of '1' outcomes for the qubits involved in the corresponding Pauli term. 
The final expectation value is computed as $\langle E \rangle = \sum_i c_i \langle\psi|P_i|\psi\rangle$. The accuracy of observable computation depends on allocating sufficient shots to achieve statistical significance while balancing execution efficiency.

Evaluating the performance of observable computation is more complex than basic circuit fidelity assessment. Each observable computation requires that the circuit under test be executed multiple times, with different basis rotations appended to measure the quantum state in the basis corresponding to the terms of the Hamiltonian. Such a performance evaluation must consider not only the complete set of circuit executions as a unified process but also the efficacy of techniques for preparing and minimizing the overhead in producing the circuits themselves.


\subsection{Existing Methods for Comparison}
\label{subsec:existing-observable-methods}

Here, we first describe several existing methods used for computing observables.  We provide these as references when assessing the performance of the new methods introduced in this work.

\subsubsection{Classical Matrix Exponentiation}
\label{subsec:matrix-exponentiation}

Given a Hamiltonian $H$ and a starting wavefunction $\ket{\psi_0}$, the state at different times $t$ is $\ket{\psi(t)}=\exp(-iHt)\ket{\psi(t=0)}$. It can be computed classically for small system sizes. Usually, it is desired to obtain the state at equally spaced time intervals. Here, we use the \texttt{expm\_multiply} routine from the Sparse Linear Algebra library from SciPy to implement this.

\subsubsection{Qiskit Estimator Primitive}
\label{subsec:estimator-primitives}

The Qiskit Estimator~\cite{qiskit_estimator_v2} provides a high-level interface for computing expectation values of quantum observables, particularly Hamiltonians, on quantum hardware or simulators. Using its \texttt{run()} method, a Qiskit circuit defining a quantum state is submitted, along with a set of observables used to estimate expectation values. The Estimator manages the underlying process of creating the necessary measurement circuits and processing results automatically. It employs optimization techniques from Qiskit's core modules, including commuting term grouping, measurement basis transformations, adaptive shot allocation to reduce circuit execution overhead, and error mitigation~\cite{Kim_2023, murali2019noiseadaptivecompilermappingsnoisy}. By streamlining the observable computation, the Estimator is a useful component in a Qiskit implementation of variational algorithms and quantum simulations.

We use the Qiskit Estimator as a helpful reference for state-of-the-art computation of expectation values on quantum computing systems that make use of the Qiskit SDK.  This allows us to compare our observable estimation computations with those obtained using the Qiskit Estimator primitive.

\subsubsection{CUDA Quantum Spin Operator}
\label{subsec:spin-operator}

The CUDA Quantum \texttt{SpinOperator} provides a high-level API for representing Hamiltonians and generating quantum kernels for execution on classically implemented quantum simulators or physical quantum hardware devices. It leverages NVIDIA GPUs for accelerated simulation, enabling efficient tensor contractions and large-scale quantum state evolution. Users define quantum observables as weighted sums of Pauli terms.  
The \texttt{SpinOperator} itself does not perform circuit decomposition or execution. Instead, it provides a structured representation of Hamiltonians that can be applied within CUDA Quantum kernels.
Additionally, CUDA Quantum provides hybrid quantum-classical execution capabilities, multi-processing features, and dynamic circuit construction, offering possibilities for novel optimization methods. 

We use the CUDA Quantum implementations of our simulation cases to push the limits of what is possible with GPU-accelerated simulation of quantum hardware devices.  CUDA Quantum kernels executed on such systems represent what ideally would be possible with fault-tolerant error-corrected quantum hardware devices in the future.  While these simulations are limited in terms of the number of qubits in width, there are no limits to the depth of the quantum circuits that can be executed.

\subsection{New Methods Explored Here}
\label{subsec:new-observable-methods}

For this work, we use three techniques developed and implemented within the benchmarking framework. Taken together, these can be shown to significantly improve both the execution time and the quality of the result obtained.

While these techniques may have been studied separately,  there has been no practical framework that integrates them into a unified and modular system. Our work fills this gap by incorporating and optimizing these techniques into a single implementation that can be readily applied across a wide range of use cases.

\subsubsection{$k$-commuting Groups}
\label{subsec:nqubit-commuting-groups}

In this section, we will describe the notion of $k$-commutativity \cite{dalfavero2312k}, and its uses to build efficient measurement circuits. 
Consider the following $n-$qubit Pauli strings:
\begin{align}
    &A = A_1 \otimes \dots \otimes A_n,
    &B = B_1 \otimes \dots \otimes B_n,
\end{align}
where each $A_i$ and $B_i$ are either a Pauli matrix or the $2 \times 2$ identity matrix. For a
$k \in \mathbb{N}$, these two Pauli strings are called $k$-commuting if and only if $k-$qubit segments of $A$ and $B$ commute, i.e.
\begin{align}
\begin{split}
    [A_{ik} \otimes \dots \otimes A_{(i+1)k}, \:\:B_{ik} \otimes \dots \otimes B_{(i+1)k} ] = 0,
\end{split}
\end{align}
for every integer $i < n/k$.

A $k-$commuting group of Pauli strings can be simultaneously diagonalized by tensor product of $k-$qubit Clifford unitaries.  As an example, consider the following Pauli strings:
\begin{align}\label{eq:2commutative}
    \{\mathrm{XXII,\:YYZZ,\:ZZXX}\}.
\end{align}
These Pauli strings are 2-commutative: $\mathrm{XX,\:YY}$ and $\mathrm{ZZ}$ commute with each other, and $\mathrm{II,\:ZZ}$ and $\mathrm{XX}$ commute with each other.
If $U_1$ simultaneously diagonalizes $\mathrm{XX,\:YY, \:ZZ}$, and $U_2$ simultaneously diagonalizes $\mathrm{XX,\:ZZ}$, 
then $U = U_1 \otimes U_2$ simultaneously diagonalizes the Pauli strings in \eqref{eq:2commutative}.

It has been shown in~\cite{dalfavero2312k} that for many Hamiltonians, as $k$ increases, the number of Pauli groups initially decreases and then saturates. In many cases, the number of groups for $k$-commuting grouping becomes equivalent to that of fully commuting grouping even when $k<n$. This makes $k$-commuting grouping preferable, as it enables shorter Clifford diagonalization circuits. Smaller values of $k$ generally result in reduced circuit depth. Instead of constructing a diagonalization circuit over all $n$ qubits, we can build shorter circuits for subsets of $k$ qubits and combine them, leading to a shorter overall measurement circuit. From a practical perspective, when quantum hardware is subject to significant noise, reducing the circuit depth becomes critical.
In scenarios where the hardware is particularly noisy or the ansatz circuit is relatively shallow, such that the depth of the measurement circuit becomes a limiting factor,
it may be necessary to choose a smaller $k$ to reduce the depth of the measurement circuit, although this may require more circuit executions and shots. Therefore, we focus on $k$-commutativity rather than using full commutativity directly. 

The grouping procedure is performed using the sorted insertion method~\cite{crawford2021efficient}. It begins by sorting the Pauli terms of the Hamiltonian based on the absoulte value of their coefficients. Then the terms are grouped greedily: each Pauli term is inserted into the first existing group that satisfies the commutativity requirement; if no such group exists, a new group is created. The initial sorting then leads to  groups with coefficients of similar magnitude, which will be important when assigning number of shots to the measurement circuits in future sections.

\subsubsection{Diagonalization of k-Commuting Groups}
\label{subsec:measurement-circuit-diagonalization}

After we group the Pauli strings into $k-$commuting groups, we then simultaneously diagonalize the Pauli strings in each group. Since each term is a Pauli string, and each group consists of $k-$commuting elements, the diagonalization can be performed using a tensor product of $k-$qubit Clifford circuits where each of them simultaneously diagonalizes a $k-$qubit segment of each Pauli string in the group. There are efficient methods for finding a Clifford circuit that diagonalizes a set of commuting Pauli strings \cite{aaronson2004improved, crawford2021efficient, kissinger2019pyzx, schmitz2024graph, bravyi2021clifford}. Here, we present a method inspired by the implementation of Pauli string rotation circuits.

Let us define the \textit{minimum support} of a Pauli string $A = A_1 \otimes A_2 \otimes \dots \otimes A_k$ as the minimum qubit index $i$ such that $A_i \neq I$. For example, minimum support of $\mathrm{XYII}$ is $1$, $\mathrm{IZIY}$ is $2$ and $\mathrm{IIZX}$ is $3$. To diagonalize a set of $k-$qubit Pauli strings, we first sort the Pauli strings with respect to their minimum support. We then diagonalize the Pauli string with the smallest minimum support by transforming it into the $Z$ matrix acting on the qubit that corresponds to its minimum support, and $I$ on all other qubits. For example, if the Pauli string is $\mathrm{IXYIIZ}$, we transform it into $\mathrm{IZIIII}$. 
This can be done via applying the sequence of Hadamard and $S$ gates, and a CNOT ladders that is used to build Pauli string rotation circuits \cite{nielsen2010quantum,kokcu2022fixed,magoulas2023cnot}. We then apply these Clifford circuits to the rest of the Pauli strings efficiently \cite{nielsen2010quantum}, and obtain a new set of Pauli strings that has one less element. Among these new Pauli strings, we pick the one with the minimum support, and apply the same procedure all over in an iterative manner until the entire set is diagonalized. 

\subsubsection{Weighted Shot Distribution}
\label{subsec:weighted-shot-distribution}

Here, we discuss our approach designed to distribute the total number of available shots such that term groups with higher coefficients receive more shots than those with smaller coefficients. This ensures that higher-weighted terms, which contribute more significantly to the computed expectation value, are measured with greater precision, leading to a higher result quality with fewer total shots. The user can adjust the tradeoff between precision and efficiency. In our work, we quantify the performance gains observed using this approach.  

To measure an operator expectation value $\left<{H}\right> = \sum_{i=1}^L c_i \left<{P_i}\right>$, we first group the Pauli strings via the $k-$commutative grouping. Let us denote the number of groups by $N$. Relabeling the coefficients and the Paulis as $c_{ij}$ and $P_{ij}$ where $i$ is the label of the group, and $j$ is the label of the term in the group, we then have $\left<{H}\right> = \sum_{i=1}^N \sum_j c_{ij} \left<{P_{ij}}\right>$. Each group can be measured with the same circuit. To obtain a precision of $\epsilon$ in the total measurement, we aim at $\epsilon/N$ precision for each $k-$commuting group, which can be satisfied by setting the shot count per group as
\begin{align}\label{eq:precision_shot_allocation}
    g_i = \frac{ N^2 \mathrm{max}_j \{ c_{ij}^2\}}{\epsilon^2}.
\end{align}
 For a given number of shots, this corresponds to allocating a different number of shots to each group of Paulis in proportion to the square of the maximum weight of the entire group.
Therefore, for a total available shot budget \( S_{\text{total}} \), we compute an initial shot allocation for each circuit as:  
\begin{equation}
s_i = \operatorname{round} \left( S_{\text{total}}  \frac{v_i}{\sum_{j=1}^{N} v_{j}} \right),
\end{equation}
where $v_i$ can be set to the squared maximum weight of the group, corresponding to the precision calculation given in \eqref{eq:precision_shot_allocation}. We also set $v_i$ to the maximum, mean, and mean squared absolute value of the weight in the entire group, and some of them turn out to be more advantageous for certain Hamiltonians.
In cases where the Hamiltonian has highly imbalanced weights, i.e., some groups contain Pauli terms with significantly larger coefficients while others consist of terms with much smaller ones, this shot allocation may result in $s_i = 0$ for certain groups, meaning that the corresponding circuit is not executed. Simulation results show that skipping circuits in this way dramatically decreases the result quality.  Therefore, we assign at least one shot to every circuit, setting $s_i = 1$ if $s_i = 0$, thereby ensuring all groups are measured.

Since rounding introduces small errors, the total number of allocated shots may not exactly match the available shot budget \( S_{\text{total}} \). To correct this, we adjust the largest allocated shot count by updating:
\begin{equation}
s_{\max} = s_{\max} + \left( S_{\text{total}} - \sum_{i} s_i \right),
\end{equation}
where $s_{\text{max}}$ is the maximum value among the initial rounded allocations $\{s_1, s_2, ...s_N \}$. This ensures that the final shot allocation remains consistent with the total available shots.

A key consideration is the balance between precision and execution efficiency. 
Suppose shots are distributed equally across all circuits. In that case, they can be submitted as a single job as an array of \( N \) circuits, where $N$ is the number of groups, minimizing initialization time between executions.
However, if each circuit is assigned a different number of shots, then each circuit must be executed individually. This results in significantly longer execution times due to the per-execution initialization overhead required by quantum hardware.  To mitigate this, we \textit{bin circuits} with similar shot counts into a reduced number of execution groups \( B_k \), such that the number of bins does not exceed a predefined fraction of the total number of circuits \( N \) (e.g., 20\%). Each bin $B_k$ is assigned a single representative shot count $S_k$, computed as the average of the original shot counts of the circuits $C_i$ within that bin:  
\begin{equation}
S_k = \frac{1}{|B_k|} \sum_{C_i \in B_k} s_i.
\end{equation}
This reduces the number of distinct executions, optimizing hardware utilization while preserving the relative importance of different term groups.   

By balancing execution overhead with measurement precision, this approach minimizes the total number of unique executions required while still ensuring that the most important Pauli terms are computed with sufficient accuracy. As a result, we can achieve the same level of result quality with fewer total shots, making quantum computations more efficient in practice.

\section{Results and Analysis}
\label{sec:results-analysis}

Here, we present the results of executing the performance assessment benchmark programs using various models from HamLib, including condensed matter physics and chemistry. We identify the differences in performance that result from using multiple optimization options, including grouping Pauli terms based on $k$-commutativity, generation of Clifford measurement circuits, and weighted shot distributions for circuit generation and observable calculation. 

The framework for assessing the performance of quantum simulations can be executed in either the Qiskit or a CUDA Quantum environment, each supporting different capabilities.
To make the code usable and easy to maintain, we have structured the code so that the problem setup, the sequencing of various test procedures, and the collection of metrics and plotting of results are independent of the lower-level code that uses these APIs to construct and execute the circuits or kernels used within the simulation.

\subsection{Experimental Setup}

\textbf{Evaluation setting:} There are two experimental settings. First, our framework focuses on evaluating the impact of Pauli string grouping methods and weighted shot distribution techniques on the accuracy of expectation value estimation. To isolate these effects, we set the Trotter step to zero, thereby eliminating Trotterization error.  We evaluate the Hamiltonian using randomly generated initial states. Each initial state is prepared using \texttt{EfficientSU2} from Qiskit with randomly sampled parameters. We repeat the evaluation 100 times with different parameter sets to account for statistical variability. Second, we evaluate execution time on both simulators and quantum hardware, using one Trotter step with a fixed initial state to perform end-to-end simulations.

\textbf{Benchmarks:} The benchmarks are divided into two categories: models from condensed matter physics, including transverse-field Ising model (TFIM), Heisenberg, Fermi-Hubbard, and Bose Hubbard models, and chemical systems, including H$_2$, B$_2$, NH, and CH molecules.  The detailed properties of these Hamiltonians, obtained from HamLib, are summarized in Table~\ref{tab:1}.

\begin{table}[h!]
\centering

\caption{Properties of Hamiltonians obtained from HamLib.}

\begin{tabular}{c|c|c}

\hline
\textbf{Models} & \textbf{Number of qubits} & \textbf{Number of terms in $H$} \\ \hline
TFIM             & 12                       & 24                          \\ 
Heisenberg       & 12                      & 48                           \\ 
Fermi-Hubbard    & 12                     & 43                           \\ 
Bose Hubbard     & 12                       & 179                          \\ 
H$_2$            & 12                       & 327                            \\ 
B$_2$            & 12                       & 287                            \\ 
NH               & 12                       & 631                            \\ 
CH               & 12                       & 631  \\                         
\hline
\end{tabular}

\label{tab:1}
\end{table}

\textbf{Metrics:} In the first experimental setting, we evaluate the fidelity improvements provided by the algorithmic optimization methods by measuring the error between simulated and theoretical energies. Theoretical energies are computed by evolving the initial states under the target Hamiltonian using NumPy. We then execute the quantum circuits using the Qiskit Aer Simulator and compare the results with the theoretical values. The error is quantified as the standard deviation of the differences between simulated and theoretical energies across multiple runs. In the second experimental setting, which focuses on evaluating execution time and result quality in end-to-end simulations, we report both the total execution time and the computed expectation values.

\subsection{Impact of Grouping Methods}

\begin{figure}
    \centering
    \includegraphics[width=\linewidth]{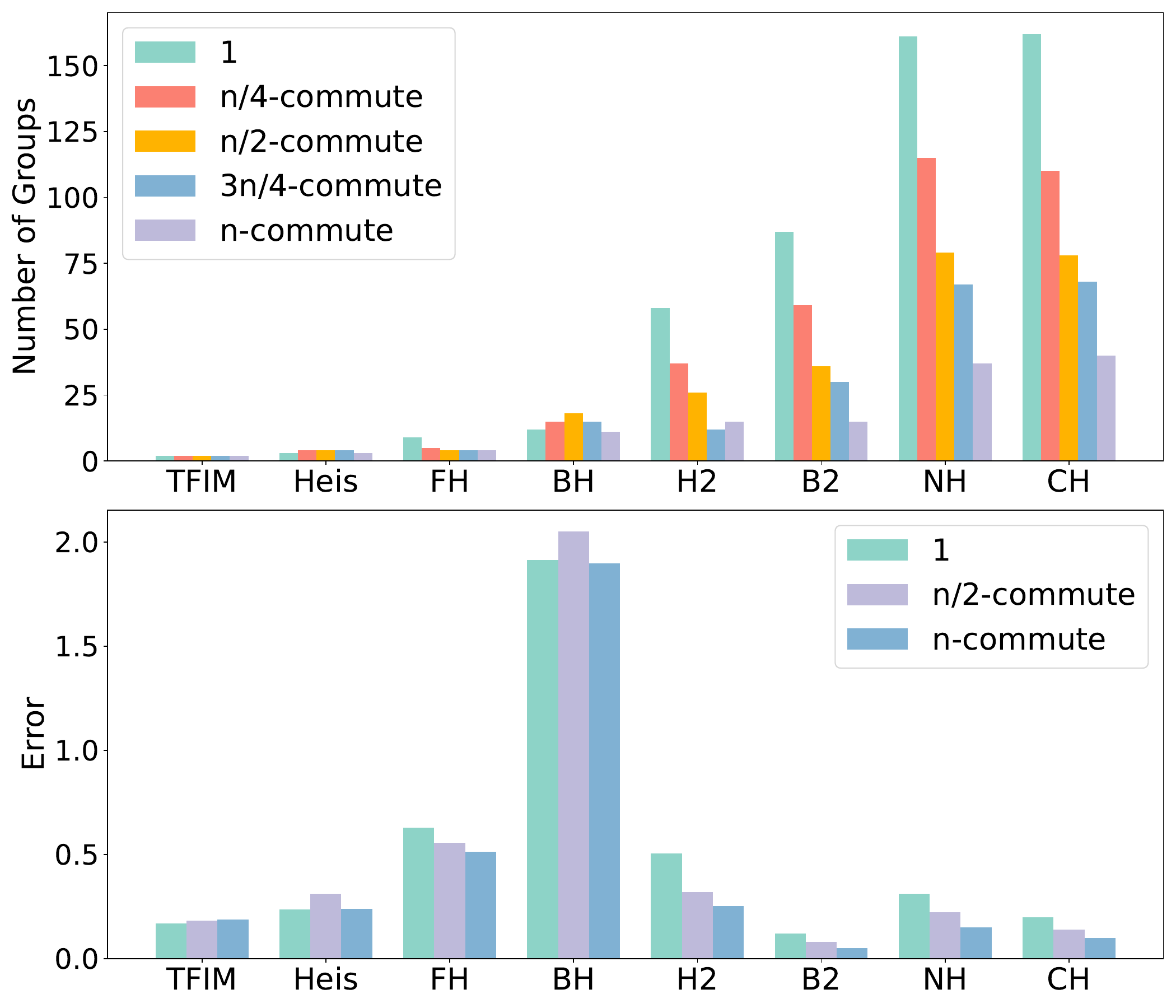}
    \caption{\textbf{Number of Groups and Expectation Compute Errors with $k$-commuting Method.} $k$ is set to 1, $\frac{n}{4}, \frac{n}{2}, \frac{3n}{4}$ and $n$. Generally, as $k$ increases, both the number of groups and errors decrease. }
    \label{fig:exp-1}
\end{figure}

We first group the Pauli strings using $k$-commutativity methods by setting $k$ equal to $1, \frac{n}{4}, \frac{n}{2}, \frac{3n}{4},$ and $n$ where $n$ is the total number of qubits in the Hamiltonian. The results are shown in ~\autoref{fig:exp-1}. For all the chemistry models, the number of groups significantly decreases as we increase $k$. Meanwhile, for condensed matter physics models, the performance of $k$-grouping is less stable. In certain cases, such as the Fermi-Hubbard model, increasing $k$ leads to the same number of groups; in this case, a smaller $k$ is preferable to reduce the depth of the measurement circuit. 
Theoretically, increasing $k$ should maintain or reduce the number of groups, as higher $k$ implies greater commuting flexibility. However, our grouping algorithm employs a greedy approach, assigning each new Pauli term to a new or an existing commuting group without guaranteeing an optimal grouping solution for the smallest number of groups. This can occasionally result in an increased total number of groups, particularly for condensed matter physics models such as the Heisenberg and Bose Hubbard models in ~\autoref{fig:exp-1}. Despite these fluctuations, the overall trend demonstrates a decrease in the number of groups as $k$ increases. 

Given that larger $k$ typically yields fewer groups, we evaluate expectation value computation errors using $k$ values of $1, \frac{n}{2},$ and $n$. The results, shown in~\autoref{fig:exp-1}, indicate that larger $k$ values generally produce fewer groups and subsequently lower errors, achieving an average error reduction of 27.1\% compared to the single-qubit commuting method.

It is important to note that the computed expectation errors strongly depend on the coefficients of the Hamiltonian. Since our primary focus is comparing different algorithmic optimization techniques, the errors reported here are not normalized by the Hamiltonian norm. We speculate that is why Hamiltonians with a larger number of terms can exhibit smaller computed errors such as in the chemistry models.

\subsection{Impact of Weighted Shot Distribution Methods}
\label{sec:weighted_shot_distribution}

\begin{figure}
    \centering
    \includegraphics[width=\linewidth]{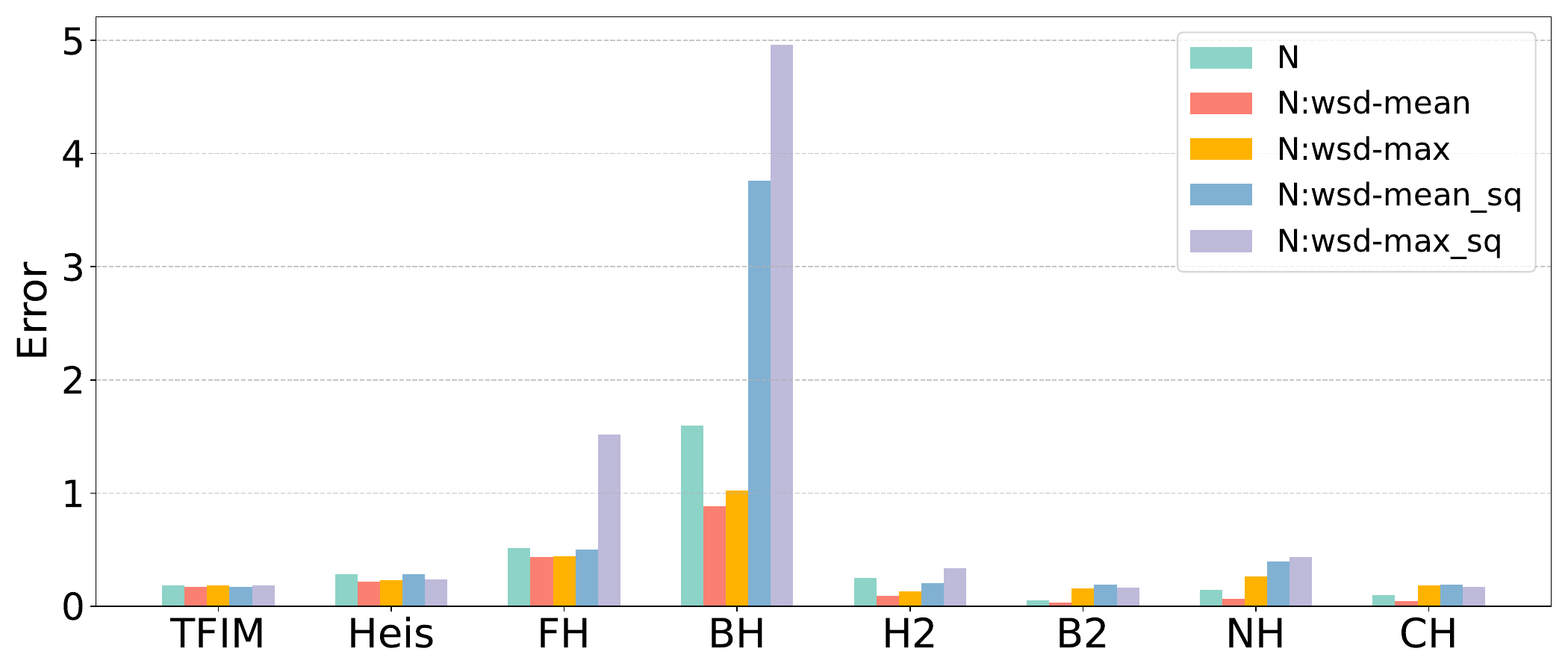}
    \caption{\textbf{Expectation Compute Errors with Uniform and Weighted Shot Distribution Methods.} The mean-weighted shot distribution method (N:wsd-mean) achieves the lowest errors and significantly outperforms the uniform shot distribution (N). In contrast, the max-squared weighted (N:wsd-max\_sq) and mean-squared weighted (N:wsd-mean\_sq) shot distribution methods lead to higher errors compared to the uniform approach. These results highlight the importance of selecting an appropriate weighted distribution strategy.}
    \label{fig:exp2-a}
\end{figure}

Based on the evaluations of the impact of different grouping methods, the fully commuting grouping technique yields the fewest groups and the lowest errors in expectation values. Therefore, we fix the grouping strategy to the $n$-commute method and focus on evaluating various weighted shot distribution strategies. Each group corresponds to one circuit, and it is allocated to a certain number of shots based on the maximum, maximum squared, mean, and mean squared absolute value of the Pauli coefficients within the group. We compare the weighted shot distribution methods against a uniform shot distribution baseline, with the total number of shots fixed at 4,000. The resulting errors in the expectation values are shown in~\autoref{fig:exp2-a}. Among all the weighted strategies, the mean-weight distribution method consistently achieves the lowest errors, demonstrating significant improvement over uniform distribution by 37.6\%. Notably, it yields a maximum improvement of 62.3\% for H$_2$. In contrast, the performance of other weighted methods is less stable and varies across different cases. In our framework, the shot distribution strategy is configurable, allowing users to select from various statistical weighting methods based on their needs.

\begin{figure}
    \centering
    \includegraphics[width=\linewidth]{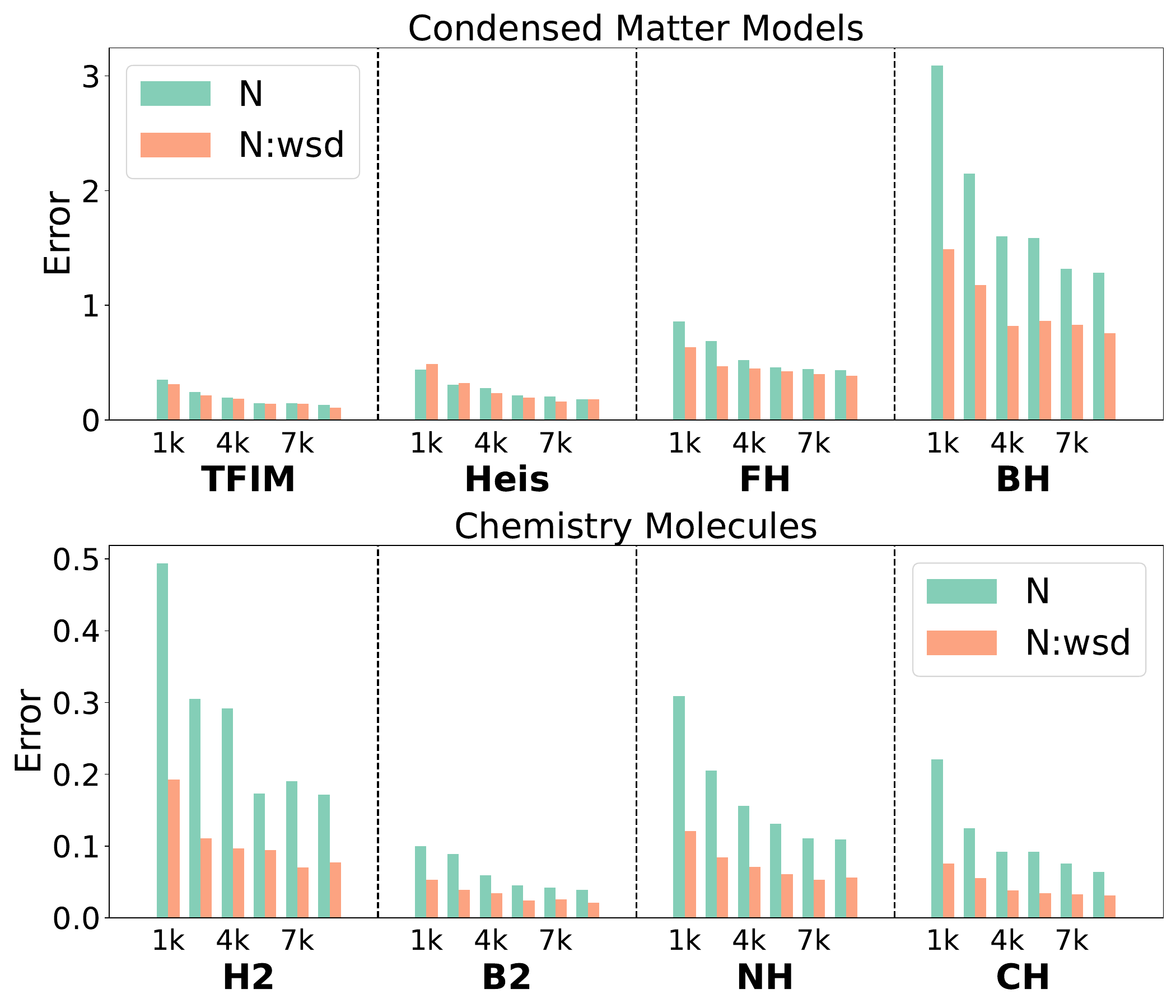}
    \caption{\textbf{Expectation Compute Errors with Varying Numbers of Shots for Uniform and Weighted Shot Distribution Methods.} The number of shots ranges from 1,000 to 8,500 in increments of 1,500. As the number of shots increases, the expectation compute errors decrease. Notably, the mean-weighted shot distribution method (N:wsd) achieves the same level of accuracy as the uniform shot distribution (N) while using significantly fewer shots. This demonstrates a substantial reduction in required shots which leads to a considerable speedup in achieving a given precision.}
    \label{fig:exp2-b}
\end{figure}

Moreover, we select the mean-weighted shot distribution method, which demonstrated the best overall performance, and compare it with uniform distribution across different shot counts. The number of shots is varied from 1,000 to 8,500 in increments of 1,500. \autoref{fig:exp2-b} shows the results. Assuming a baseline of 8,500 shots under uniform distribution, the mean-weighted method achieves the same level of error precision with over 50.7\% fewer shots on average, corresponding to an average 2.6× speedup in evaluation time. In several cases, including Bose Hubbard, H$_2$, B$_2$, NH, and CH, the uniform distribution requires 8,500 shots to match the accuracy achieved by the mean-weighted method with fewer than 2,500 shots, reflecting a more than 3.4× reduction in shot overhead for achieving equivalent precision.

\subsection{Scaling Challenges: Execution Time and Result Quality}
\label{sec:scaling_challenges}

In this section, we evaluate how the execution time and result quality are impacted as the simulation problem scales to use larger numbers of qubits.
In the previous sections, we examined the performance of several new methods for expectation value computation. Here, we broaden the scope to include end-to-end simulation performance characteristics, presenting execution time trends as a function of the number of qubits along with the expectation values observed at each size. Our results highlight the full impact that the choice of observable computation method has on the simulation's scaling characteristics.

For these experiments, we implement one step of a Hamiltonian evolution from a known initial state, which is the Neel state, over problem sizes ranging from 4 to 42 qubits. The Hamiltonian corresponds to the TFIM, evolved using a single-step Trotterized unitary at short time t = 0.1, with 10,000 shots and a single repetition per run. Computation of the expectation value is performed using the methods described in sections~\ref{subsec:existing-observable-methods} and~\ref{subsec:new-observable-methods}.

\begin{figure}[t!]
\includegraphics[width=1.0\columnwidth]{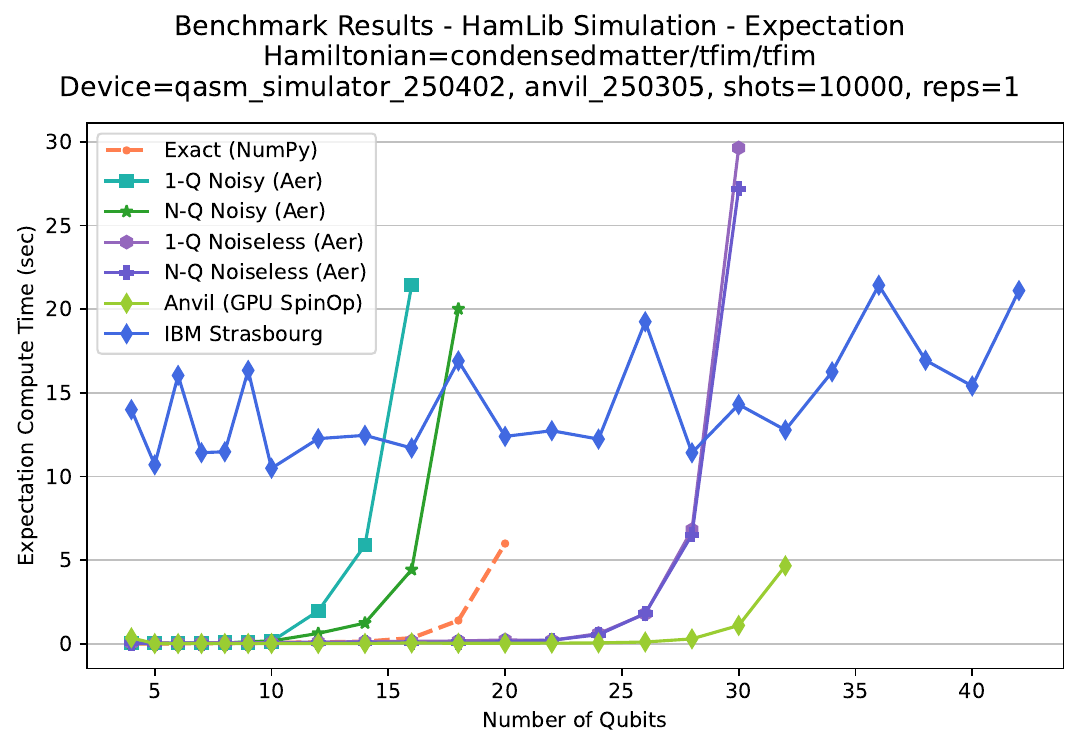}
\caption{
\textbf{Execution Time Scaling for Expectation Computation in TFIM Simulations.}
This plot illustrates the total compute time required to evaluate the expectation value of the TFIM Hamiltonian using a variety of classical and quantum simulation methods. Techniques include exact diagonalization, Aer-based simulations with and without noise, GPU-accelerated simulation on Purdue’s Anvil system, and execution on the IBM Strasbourg quantum hardware. The results demonstrate a wide range in scalability and performance, with the classical GPU method offering the fastest runtime, while quantum hardware shows modest growth in time, even at high qubit counts. The variability in IBM hardware timing is likely influenced by queuing dynamics and system-level fluctuations.
}
\label{fig:simulation_time_with_hardware}
\end{figure}

\vspace{0.2cm}

\autoref{fig:simulation_time_with_hardware} presents the results from executing this Hamiltonian simulation using a variety of simulation strategies with a focus on the total execution time. The x-axis denotes the number of qubits simulated, ranging from 4 to 42, while the y-axis shows the corresponding execution times in seconds. Our evaluation spans several approaches: exact diagonalization using NumPy, noise-free and noise-injected simulations via the Qiskit Aer backend, GPU-accelerated simulation on Purdue’s Anvil cluster using NVIDIA hardware, and execution on the IBM Strasbourg quantum processor. The labels “1-Q” and “N-Q” denote grouping strategies for commuting observables, corresponding to $k=1$ commuting group and $k=n$ commuting group, respectively. The N-Q commuting group algorithm produces groups with similar magnitude coefficients, which enables concomitant use of the weighted shot distribution algorithm.
The Aer simulations were performed using 10,000 shots, with the exception of the N-Q cases, which used 2,500 shots to illustrate the benefit of weighted shot distribution discussed in section~\ref{sec:weighted_shot_distribution}.

The “N-Q Noisy” trace shows a fourfold reduction in execution time compared to the “1-Q Noisy” case, owing directly to the reduced shot count — an anticipated advantage of measuring larger commuting groups without compromising result fidelity. In contrast, the difference between 1-Q and N-Q in the noiseless Aer simulations is negligible. This is because, in the absence of noise, the Aer backend employs an optimized statevector simulation that synthesizes measurement outcomes efficiently, making the shot count largely inconsequential.

The Anvil (GPU) results underscore the power of classical high-performance computing: by leveraging matrix-based simulation, it demonstrates exceptionally fast runtimes and establishes a useful performance baseline against which quantum hardware can be compared.
This simulation - running on a state-of-the-art GPU and using the CUDA Quantum SpinOperator class - consistently outperforms others in speed, highlighting the advantage of GPU acceleration for quantum simulation workloads.
This benchmark serves as a reference point that illustrates the limits of GPU simulation performance. In future work, we plan to explore the use of advanced sampling techniques and improved grouping algorithms on this platform, which could reduce execution times and enable even larger simulations.

\begin{figure}[t!]
\includegraphics[width=1.0\columnwidth]{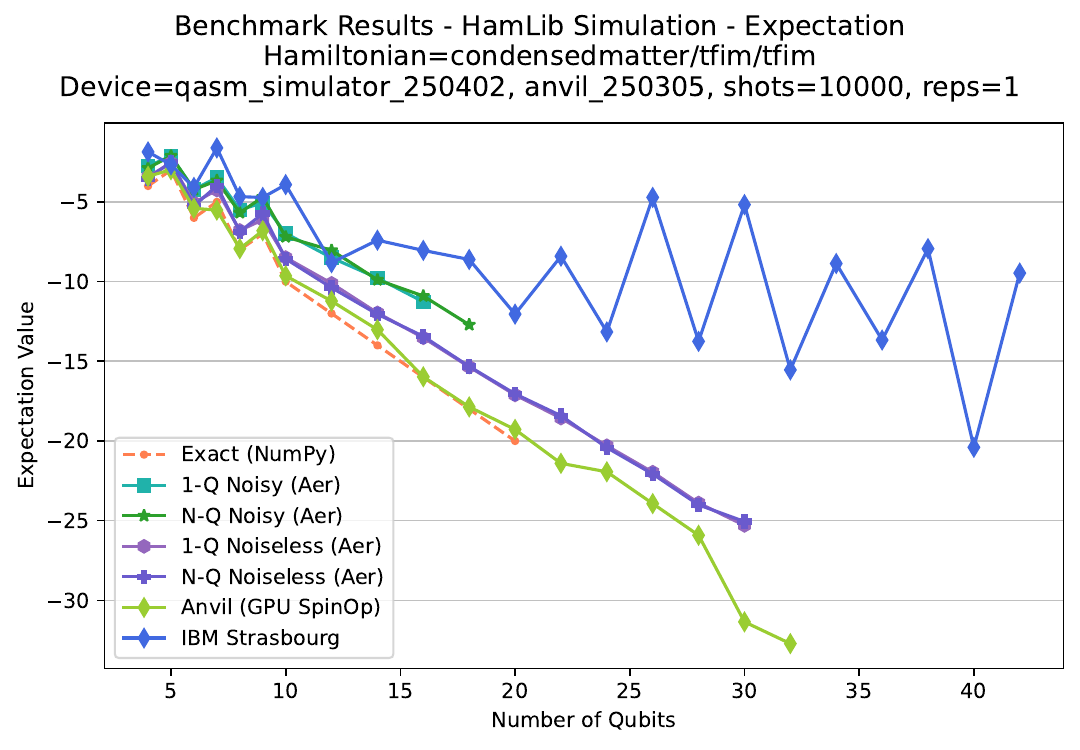}
\caption{
\textbf{Accuracy of Expectation Values in TFIM Simulations Across Platforms.}
This plot presents the computed expectation values for the TFIM Hamiltonian across increasing qubit counts. The “Exact” values, calculated via matrix exponentiation, serve as a noise and Trotter-error-free reference. Other methods exhibit deviations due to Trotterization, noise, or both. Notably, the IBM hardware results diverge significantly beyond 12 qubits, likely due to increased noise and limited shot resolution. GPU-based and noiseless simulations show closer agreement with the exact values, while noisy Aer and hardware runs illustrate the tradeoffs in fidelity that accompany real-device execution. These results highlight the impact of error sources on simulation accuracy and underscore the importance of choosing appropriate methods for large-scale Hamiltonian simulation. 
}
\label{fig:simulation_value_with_hardware}
\end{figure}

\vspace{0.2cm}

\autoref{fig:simulation_value_with_hardware} shows the computed expectation values for the TFIM simulation across qubit counts ranging from 4 to 42, corresponding to the same experiments shown in~\autoref{fig:simulation_time_with_hardware}. Each data point corresponds to the result of executing a Trotterized quantum circuit with a single step (K = 1) of duration t = 0.1, followed by basis rotation and measurement. The choice of K = 1 allows the benchmark to incorporate Trotterization effects while maintaining a shallow circuit depth representative of realistic near-term quantum workloads. All simulations were performed using 10,000 shots and a single repetition, with the exception of the N-Q cases, which used 2,500 shots.

The orange dashed “Exact” line, computed using NumPy through full matrix exponentiation, serves as a reference with zero Trotter error. In comparison, both the 1-Q and N-Q Noiseless traces exhibit small but noticeable deviations from the exact values, reflecting the Trotter errors introduced by the circuit-based approximation of time evolution. These deviations become more pronounced in the 1-Q and N-Q Noisy simulations, which are subject to both Trotterization and simulated hardware noise calibrated to mimic a system with a quantum volume of approximately 2048.

The Anvil (GPU) trace also reflects the presence of Trotter error, though its deviation from the exact values is consistently smaller than that of the Aer-based noisy simulations. We hypothesize that this improvement stems from the SpinOperator formalism employed in the Anvil simulation, which may reorder the Hamiltonian terms in a manner that coincidentally reduces Trotter error. However, this observation remains speculative and warrants further investigation.

\vspace{0.2cm}

Finally, the IBM Strasbourg hardware results display considerable variability in execution time and expectation values. The following subsection provides a more detailed analysis of these results.

\subsection{Quantum Hardware Results}

The IBM Strasbourg hardware results presented in~\autoref{fig:simulation_time_with_hardware} (execution time) and ~\autoref{fig:simulation_value_with_hardware} (expectation value) stand in contrast to the behavior observed in classical and simulated approaches. These hardware results were obtained without any error mitigation techniques, as such methods impose significant overhead for large problem or circuit sizes. Notably, the growth in execution time on hardware increases much more gradually with qubit count than in any of the other simulation methods. We successfully executed circuits with up to 42 qubits — well beyond the practical limits of statevector or matrix-based simulators—while observing only a modest rise in execution time. Understanding how hardware execution time scales with Hamiltonian size and circuit complexity remains an open question and is a key direction for future investigation.

Significant variability is also evident in the hardware execution time results. We hypothesize that the observed spikes are artifacts of the batch execution mode that we employed in Qiskit Runtime. In this mode, circuits are submitted in groups; once the initial job in a batch clears the queue, subsequent circuits within the batch can execute in rapid succession. However, job scheduling by other users between batches can introduce irregular delays, resulting in the timing fluctuations observed. Alternatively, these variations may arise from instabilities in the quantum control systems themselves. While this behavior has not yet been systematically investigated, our findings suggest that both queuing dynamics and hardware-level noise could contribute to the observed variability.

Regarding the expectation values, the hardware results increasingly diverge from the exact reference value as the qubit count exceeds 12. This growing discrepancy reflects the combined impact of gate imperfections, readout noise, and the limited number of measurement shots. Unlike the noisy Aer simulations, which emulate a high-fidelity device with well-characterized error rates, an actual quantum hardware device typically exhibits a more variable and pronounced noise profile. Additionally, the use of 10,000 shots may be insufficient to capture accurate expectation values at larger system sizes, where statistical fluctuations have a greater effect. Increasing the number of shots may improve result fidelity, but this remains to be validated through further experimentation.

\section{Summary and Conclusion}
\label{sec:summary-and-conclusions}

In this work, we describe our development of a framework for exploring the performance of observable estimation in quantum simulations of physical systems in condensed matter and chemistry. While multiple implementation options exist for such simulations on quantum computing systems, they present critical tradeoffs between execution time and result quality that significantly impact their practical utility and cost.

We first reviewed the primary stages of quantum simulation: state initialization, quantum dynamics and time evolution, and observable measurement. We described several implementation and optimization options for each phase, including grouping commuting Pauli terms, generating Clifford-based measurement circuits, and distributing shots to the circuits based on the weights of Pauli terms, with a particular focus on evolution and observable computation. Our framework captures relevant metrics for each stage both independently and as an integrated process, revealing detailed insights about execution time and result quality. This systematic approach enables efficient exploration of algorithmic variations and evaluation of different implementation strategies.

The evaluation of commuting groups in our framework is primarily focused on reducing the number of required shot counts for accurate expectation value estimation. Beyond this, commuting groups could be used for the construction of time evolution circuits. Within a group, there is no error that results from Trotterization, and therefore group-based circuit construction may reduce the overall Trotter error. Grouping may also lead to optimized execution times on hardware supporting parallel operations. It is also possible that randomization of the sequence of terms during Trotterization may reduce systematic error. We leave the exploration of this with our framework to future work.

The framework serves as a development platform for users to explore their applications and investigate new algorithmic techniques. When used across different quantum computing platforms—from current NISQ devices to high-fidelity GPU-based quantum simulators—it enables systematic assessment of our progress toward fault-tolerant quantum computation. Though limited in size, the GPU simulators provide ideal quantum operations that can model how algorithms may perform in the fault-tolerant era, offering a valuable reference point for evaluating both current hardware and algorithmic approaches.

In conclusion, this study's insights refine our understanding of practical Hamiltonian simulation performance and provide a robust foundation for future research. As quantum computing continues to evolve, the strategies developed in this work will be crucial in addressing the scalability challenges of benchmarking quantum simulations.

\section*{Code Availability}
\label{sec:data_and_code_availability}

The code for the benchmark suite described in this work is available at 
\href{https://github.com/SRI-International/QC-App-Oriented-Benchmarks}{https://github.com/SRI-International/QC-App-Oriented-Benchmarks}.
Detailed instructions are provided in the repository.

\section*{Acknowledgment}
The Quantum Economic Development Consortium (QED-C), comprised of industry, government, and academic institutions with NIST support, formed a Technical Advisory Committee (TAC) to assess quantum technology standards and promote economic growth through standardization. The Standards TAC developed the Application-Oriented Performance Benchmarks for Quantum Computing as an open-source initiative with contributions from multiple QED-C quantum computing members.
We thank QED-C members for their valuable input in reviewing and enhancing this work.

We acknowledge the use of IBM Quantum services for this work. The views expressed are those of the authors and do not reflect the official policy or position of IBM or the IBM Quantum team.
IBM Quantum. https://quantum-computing.ibm.com, 2024.

We acknowledge Purdue University and the Rosen Center for Advanced Computing \cite{McCartney2014} for providing access to and execution time on their Anvil Hardware. 

EK acknowledges the support from the U.S. Department of Energy (DOE) under Contract No. DE-AC02-05CH11231, through the Office of Advanced Scientific Computing Research Accelerated Research for Quantum Computing Program.

\bibliographystyle{ieeetr}
\bibliography{references}

\end{document}